\documentclass[11pt]{cernrep}
\usepackage{amsmath,amssymb,graphicx}
\usepackage[rflt]{floatflt}
\newcommand{\bra}{\langle}
\newcommand{\ket}{\rangle}
\newcommand{\tr}{\operatorname{Tr}}
\newcommand{\psibar}{\overline{\psi}}
\newcommand{\that}{\hat{t}}
\newcommand{\kt}{\tilde{k}}
\newcommand{\Udag}{U^\dagger}
\newcommand{\eq}{\varepsilon_q}
\newcommand{\eg}{\varepsilon_g}

\begin{document}

\title{THE ONSET AND DECONFINEMENT TRANSITIONS IN TWO-COLOUR QCD}

\author{Jon-Ivar Skullerud$^1$, Simon Hands$^2$ and Seyong Kim$^{2,3}$}

\institute{$^1$School of Mathematics, Trinity College, Dublin 2, Ireland\\
$^2$Department of Physics, University of Wales Swansea, Singleton Park,
Swansea SA2 8PP, UK\\
$^3$Department of Physics, Sejong University, Gunja-Dong, Gwangjin-Gu,
Seoul 143-747, South Korea}

\maketitle

\begin{abstract}
We study two-colour QCD with two flavours of Wilson fermion at
non-zero chemical potential $\mu$ and zero temperature.  We find
evidence of two separate transitions: an onset transition at
$\mu\approx m_\pi/2$ where quark number and energy densities increase
from zero; and a deconfinement transition at higher $\mu$.  The
$\mu$-dependence of the number and energy densities and the diquark
condensate indicate that a Fermi surface is formed and that BCS rather
than Bose--Einstein condensation dominates at the quark mass
considered here, especially beyond the deconfinement transition.
\end{abstract}

\section{Introduction}

QCD with gauge group SU(2) and non-zero chemical potential $\mu$ has a
number of features that makes it an interesting object of study.
Since the quarks and antiquarks live in equivalent representations of
the colour group and can be related by an anti-unitary symmetry (the
Pauli--G\"ursey symmetry), the fermion determinant remains real at
non-zero $\mu$.  For an even number of flavours, it can also be shown
to be positive, making standard Monte Carlo sampling possible.  Thus
two-colour QCD can be used as a laboratory for investigating gauge
theories at high density and low temperature, a region of phase space
which is still inaccessible to lattice simulations of real
(three-colour) QCD.

At $\mu=0$, the Pauli--G\"ursey symmetry implies an exact symmetry
between mesons and diquarks, which are the baryons of the theory.  In
particular, chiral multiplets will contain both mesons and baryons.
For $N_f=2$ for example, the pseudo-Goldstone multiplet consists of
the pion isotriplet plus a scalar isoscalar diquark--antidiquark
pair.  The diquarks can be expected to condense when $\mu\gtrsim
m_\pi/2$, forming a superfluid ground state.  In this respect, the
theory is radically different from real QCD, where no gauge invariant
diquark condensate exist and the ground state at high density is
superconducting.  The nature of the superfluid ground state is however
an interesting issue in its own right.

When $m_\pi\ll m_\rho$ ($\rho$ denoting any non-Goldstone boson) the
$\mu$-dependence of the theory can be analysed using chiral
perturbation theory \cite{Kogut:2000ek}.  In this case a transition to
a Bose--Einstein condensate (BEC) of tightly bound diquarks occurs at
$\mu=\mu_o=m_\pi/2$, giving rise to a quark density $n_q$ and diquark
condensate $\bra qq\ket$,
\begin{equation}
n_q\propto f_\pi^2(\mu-\mu_o);\quad
\bra qq\ket\propto\sqrt{1-\left(\frac{\mu_0}{\mu}\right)^4}
\quad\Rightarrow\lim_{\mu\to\infty}\bra qq(\mu)\ket=\text{const.}
\label{eq:chiPT}
\end{equation}
This has been confirmed in simulations with staggered fermions
\cite{Aloisio:2000if,Hands:2000ei,Hands:2001ee,Kogut:2001na}.

This picture is expected to break down when $m_\pi\approx m_\rho$, at
which point the Goldstones are no longer distinguished hadrons, and
the fermionic nature of the constituents comes into play.  At this
point one may expect BCS condensation of quarks near the Fermi surface
to become the dominant mechanism.  Assuming states within thickness
$\Delta\ll\mu$ around the Fermi surface participate in the pairing, we
obtain the following qualitative behaviour:
\begin{equation}
n_q\propto \mu^3;\quad\langle qq\rangle\propto\Delta\mu^2.
\label{eq:BCS}
\end{equation}
The same rule of thumb predicts the quark energy density
$\eq\propto\mu^4$.

In the gluon sector, the differences between SU(2) and SU(3) are
expected to be less important, and 2-colour QCD is a good setting for
{\em ab initio} studies of gluodynamics in the presence of a background
baryon density.  Of particular interest is the issue of deconfinement
at high density.  Signals of deconfinement were observed in
simulations with Wilson \cite{Muroya:2002ry} and staggered fermions
\cite{Alles:2002st}, where correlations were found between the
Polyakov loop and chiral or baryonic observables.  However, the phase
structure has not been investigated in any further detail, and it
remains unclear whether there is a confined phase with non-zero baryon
number (as in QCD), or just a single phase transition.  This will be
investigated in the present paper.

\section{Lattice formulation}

The $N_f=2$ fermion action is given by
\begin{equation}
S=\bar\psi_1M(\mu)\psi_1+\bar\psi_2M(\mu)\psi_2
-J\bar\psi_1(C\gamma_5)\tau_2\bar\psi_2^{tr}
+\bar J\psi_2^{tr}(C\gamma_5)\tau_2\psi_1,\label{eq:action}
\end{equation}
where $M(\mu)$ is the usual Wilson fermion matrix
\begin{equation}
M_{xy}(\mu)=\delta_{xy}
-\kappa\sum_{\nu}[(1-\gamma_\nu)e^{\mu\delta_{\nu0}}U_\nu(x)\delta_{y,x+\hat\nu}
+(1+\gamma_\nu)e^{-\mu\delta_{\nu0}}U_\nu^\dagger(y)\delta_{y,x-\hat\nu}]\,.
\end{equation}
The diquark source terms $J,\bar{J}$ serve a double purpose in lifting
the low-lying eigenmodes in the superfluid phase, thus making the
simulation numerically tractable, and enabling us to study diquark
condensation without any ``partial quenching''.  The results will at
the end be extrapolated to the physical limit $J=\bar{J}=0$.  We will
also introduce the rescaled source strength $j\equiv J/\kappa$.

The Pauli--G\"ursey symmetry is
\begin{equation}
KM(\mu)K^{-1} = M^*(\mu)\quad\text{with}\quad K\equiv
C\gamma_5\tau_2\,.
\label{eq:PG}
\end{equation}
This symmetry implies that $\det M(\mu)$ is real, but not necessarily
positive.  However, with the change of variables $\bar\phi=-\psi_2^{tr}C\tau_2,
\phi=C^{-1}\tau_2\bar\psi_2^{tr}$ we can rewrite the action as
\begin{equation}
S=\begin{pmatrix}\bar\psi &\bar\phi\end{pmatrix}
\begin{pmatrix} M(\mu)&J\gamma_5\\
            -\bar J\gamma_5&M(-\mu)\end{pmatrix}
\begin{pmatrix}\psi\\\phi\end{pmatrix}\equiv\bar\Psi{\cal M}\Psi.
\label{eq:bilinear}
\end{equation}
It is now straightforward to show that $\det{\cal M}$ is real
and positive if $\bar{J}=J^{*}$ \cite{Skullerud:2003yc,Hands:2005yq},
and it is possible to simulate this action with a standard HMC
algorithm \cite{Hands:2005yq}.

We have computed the quark number and energy densities, diquark
condensate, gluon energy density and Polyakov loop on each
trajectory.  The quark number density is given by
\begin{equation}
n_q =
\sum_i\kappa\Bigl\bra\psibar_i(x)(\gamma_0-1)e^{\mu}U_t(x)\psi_i(x+\that)
+ \psibar_i(x)(\gamma_0+1)e^{-\mu}\Udag_t(x-\that)\psi_i(x-\that)\Bigr\ket\,.
\label{eq:nq}
\end{equation}
The energy density is given by 
\begin{equation}
\eq =
\sum_i\kappa\Bigl\bra\psibar_i(x)(\gamma_0-1)e^{\mu}U_t(x)\psi_i(x+\that)
- \psibar_i(x)(\gamma_0+1)e^{-\mu}\Udag_t(x-\that)\psi_i(x-\that)\Bigr\ket
- \eq^0\,,
\label{eq:eq}
\end{equation}
where the vacuum subtraction $\eq^0$ is
\begin{equation}
\eq^0 = \frac{4}{d}\Bigl(N_fN_c-\bra\psibar\psi\ket_{\mu=0}\Bigr)\,.
\end{equation}
The diquark condensate is 
\begin{equation}
\bra qq\ket = \frac{\kappa}{2}\bra\psibar_1K\psibar_2^{tr} -
\psi_2^{tr}K\psi_1\ket\,,
\end{equation}
while the gluon energy density is given by
$\eg=\frac{3\beta}{2}\bra\tr(P_t-P_s)\ket$, where $P_t$ and $P_s$ are
the timelike and spacelike plaquettes respectively.

\section{Results}

\begin{floatingtable}[r]{
\begin{tabular}{|l|r|r|}
\hline
$a\mu$ &$N(\ell=0.5)$ &$N(\ell=1.0)$ \\ \hline
0.3 & 300 & \\
0.35 & & 230 \\
0.4 & 300 & 90 \\
0.45 & 300 & 64 \\
0.5 & 300 & 42 \\
0.55 & 300 & \\
0.6 & 303  & 108 \\
0.65 & & 288 \\
0.7 & 302 & 48 \\
0.75 & & 137 \\
0.8 & 280 & 58 \\
0.9 & 300 &  \\
1.0 & 300 &  \\ \hline
\end{tabular}}
\caption{Run parameters for the HMC simulations.  $N$ is the
  number of trajectories, while $\ell$ is the average
  trajectory length.}
\label{tab:traj}
\end{floatingtable}
We have simulated with $\beta=1.7, \kappa=0.178$ on an $8^3\times 16$
lattice, corresponding to $a=0.22$ fm, $m_\pi/m_\rho=0.92$
\cite{Skullerud:2004int}.  We have used a diquark source $aj=0.04$
with $0.3\leq a\mu\leq1.0$.  The number of trajectories and their length
are given in table~\ref{tab:traj}.  We have also started simulations
for selected values of $\mu$ with $aj=0.02$, but those results will
not be shown here.  Configurations were saved every 4 trajectories of
length 1 and every 8 trajectories of length 0.5.

\begin{figure}
\includegraphics*[width=0.45\textwidth]{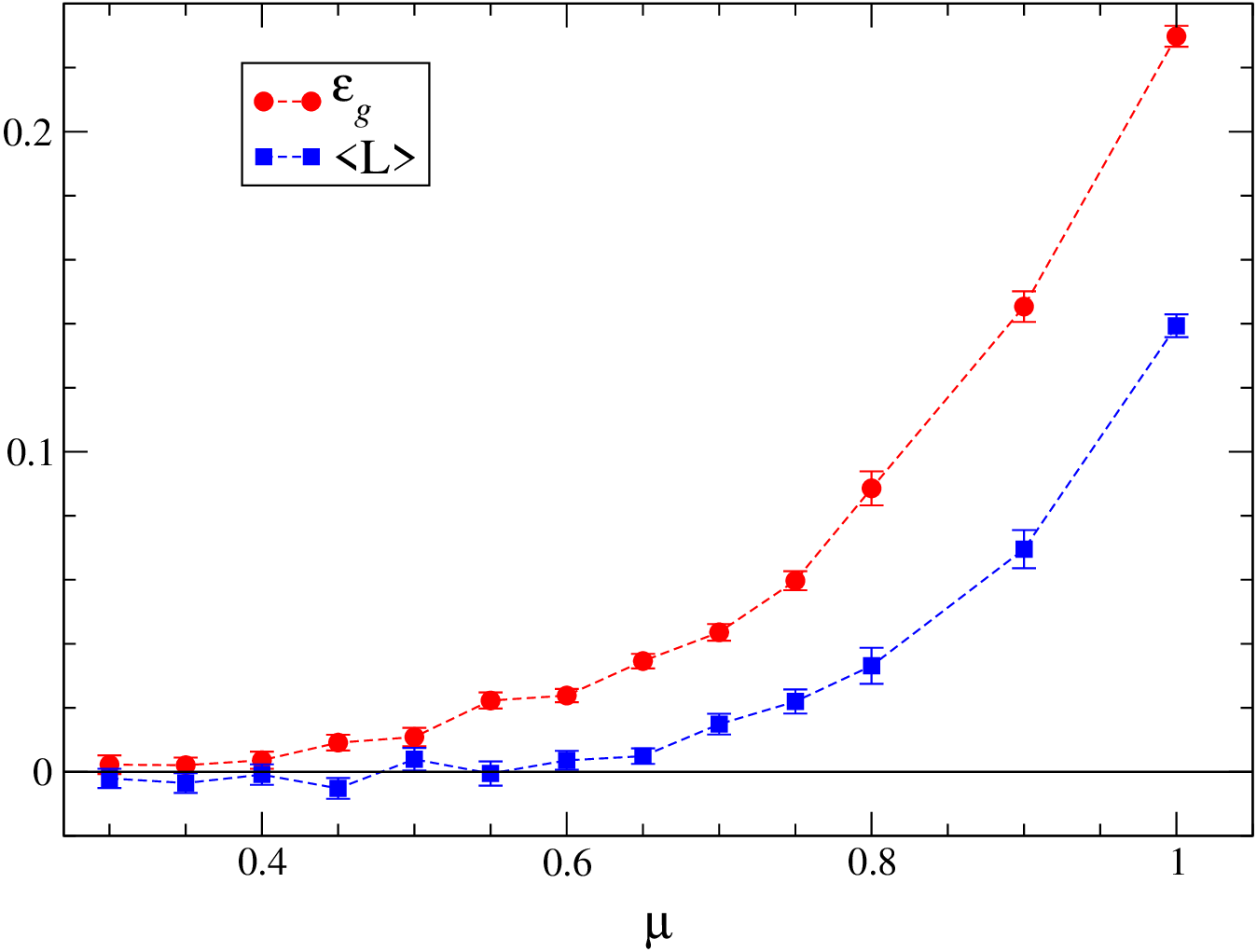}
\includegraphics*[width=0.45\textwidth]{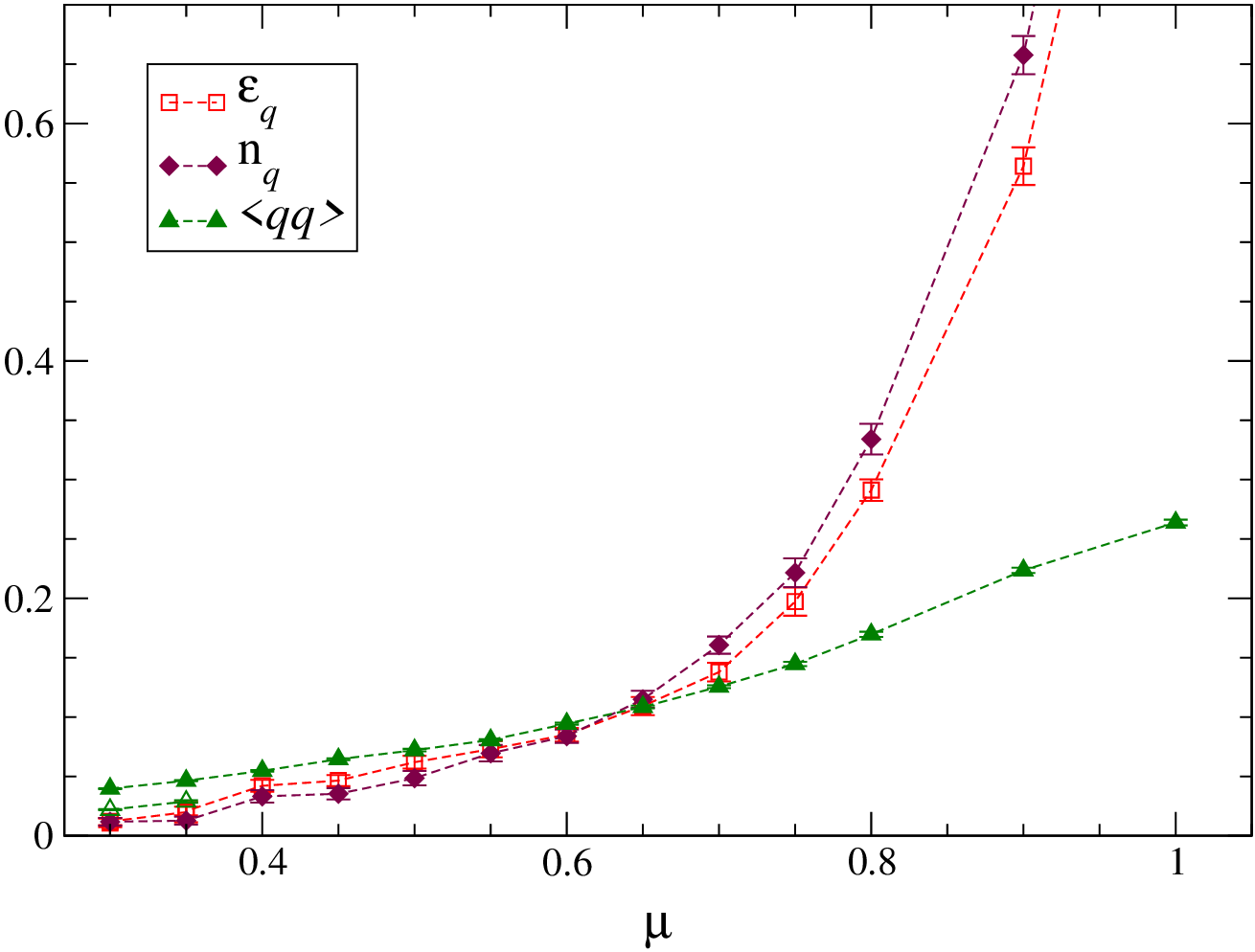}
\caption{Results for bosonic (left) and fermionic (right) quantities,
  as a function of chemical potential $\mu$, for $aj=0.04$.  The open
  triangles are preliminary results for the diquark condensate with
  $aj=0.02$.}
\label{fig:bosefermi}
\end{figure}

In figure \ref{fig:bosefermi} we show results for the Polyakov loop,
gluon and quark energy densities, quark number density and diquark
condensate, as function of chemical potential.  We see evidence of two
distinct transitions: first, an {\em onset} transition around
$a\mu=0.4$ where the quark number and energy density increase from zero;
and second, a {\em deconfinement} transition around $a\mu=0.65$ where
the Polyakov loop takes on a non-zero value.  We also find that the
chiral condensate starts to decrease from its vacuum value around
$a\mu=0.4$.

\begin{figure}
\includegraphics*[width=0.45\textwidth]{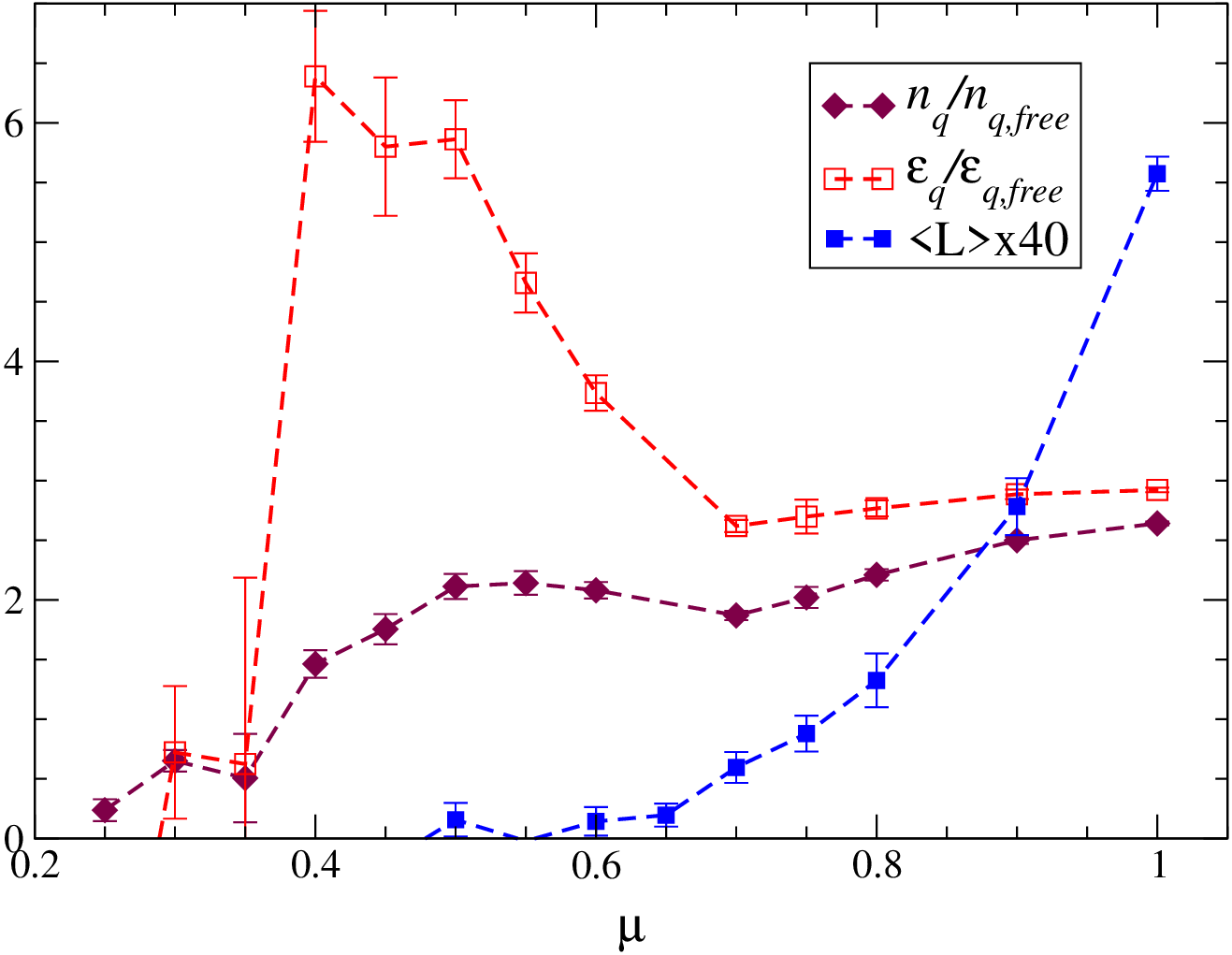}
\includegraphics*[width=0.45\textwidth]{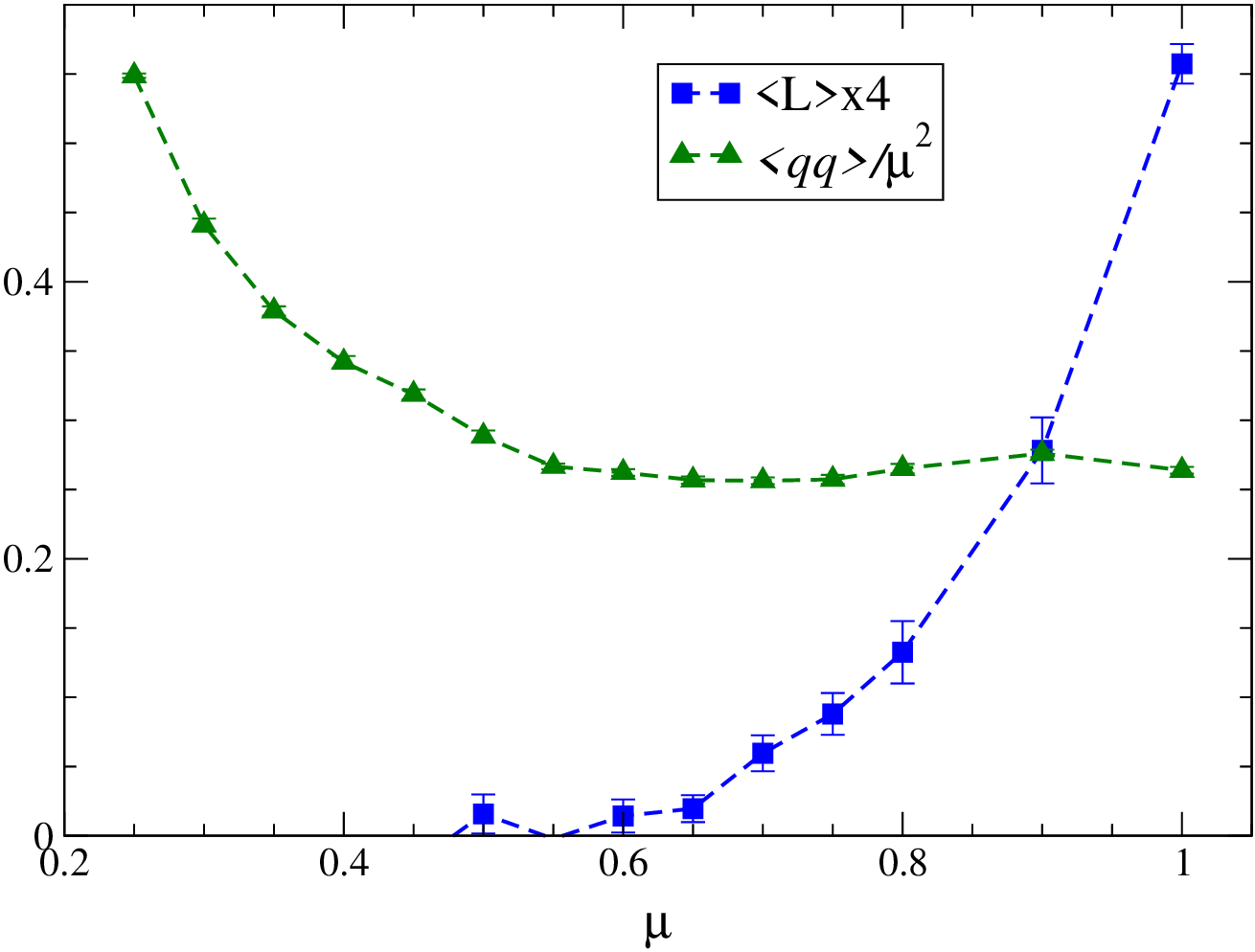}
\caption{Left: Quark number density $n_q$ and energy density $\eq$,
  divided by their values for free massless fermions on an
  $8^3\times16$ lattice.  Right: The diquark condensate $\bra qq\ket$
  divided by the BCS formula (\protect\ref{eq:BCS}), together with the
  average Polyakov loop $\bra L\ket$.}
\label{fig:rescaled}
\end{figure}

The behaviour of $n_q$ and $\bra qq\ket$ appears strikingly different
from that of (\ref{eq:chiPT}), and is suggestive of the formation of a
Fermi surface and a $\mu$-dependence qualitatively similar to that of
(\ref{eq:BCS}) \cite{Hands:2005yq}.  On the coarse lattices used here,
the continuum behaviour may however be obscured by lattice artefacts.
To compensate for this, we divide our results with the
expected behaviour of free, massless Wilson fermions on an
$N_s^3\times N_t$ lattice,
\begin{equation}
n_{q,\text{free}}=\frac{4N_fN_c}{N_s^3N_t}\sum_k{
\frac{i\sin\kt_0\bigl[\sum_i\cos k_i-4\bigr]}
{\bigl[4-\sum_\nu\cos\kt_\nu\bigr]^2+\sum_\nu\sin^2\kt_\nu}}
\label{eq:nqfree}
\end{equation}
where
\begin{equation}
\kt_\nu = 
\begin{cases}
  k_0-i\mu = \frac{2\pi}{N_t}\bigl(n_0+\frac{1}{2}\bigr)-i\mu\,, & \nu=0,\\
  k_\nu = \frac{2\pi n_\nu}{N_s}\,, & \nu=1,2,3.
\end{cases}
\end{equation}
and $N_s=8, N_t=16$ in our case.  The result of this is shown in
figure \ref{fig:rescaled}.  The number density shows good BCS scaling
already from $a\mu\approx0.5$, while no such agreement is shown for
the energy density until the deconfinement transition.  The fact that
$n_q$ exceeds the free field value by a factor of two is consistent
with a degenerate system with $k_F>E_F$, suggesting a non-zero binding
energy.  Strikingly, the gluon energy density exhibits very good
scaling with $\mu^4$ over the whole range of $\mu$ considered.

Also in fig.~\ref{fig:rescaled}, we show the diquark condensate
divided by $\mu^2$, which should be constant in the case of BCS
condensation.  Such scaling is indeed observed in the deconfined
phase, for $a\mu\gtrsim0.65$.  Below this point the behaviour appears
to be in qualitative agreement with the $\chi$PT expression
(\ref{eq:chiPT}), but this requires further study before any firm
conclusion can be drawn.

\begin{figure}
\includegraphics*[width=0.45\textwidth]{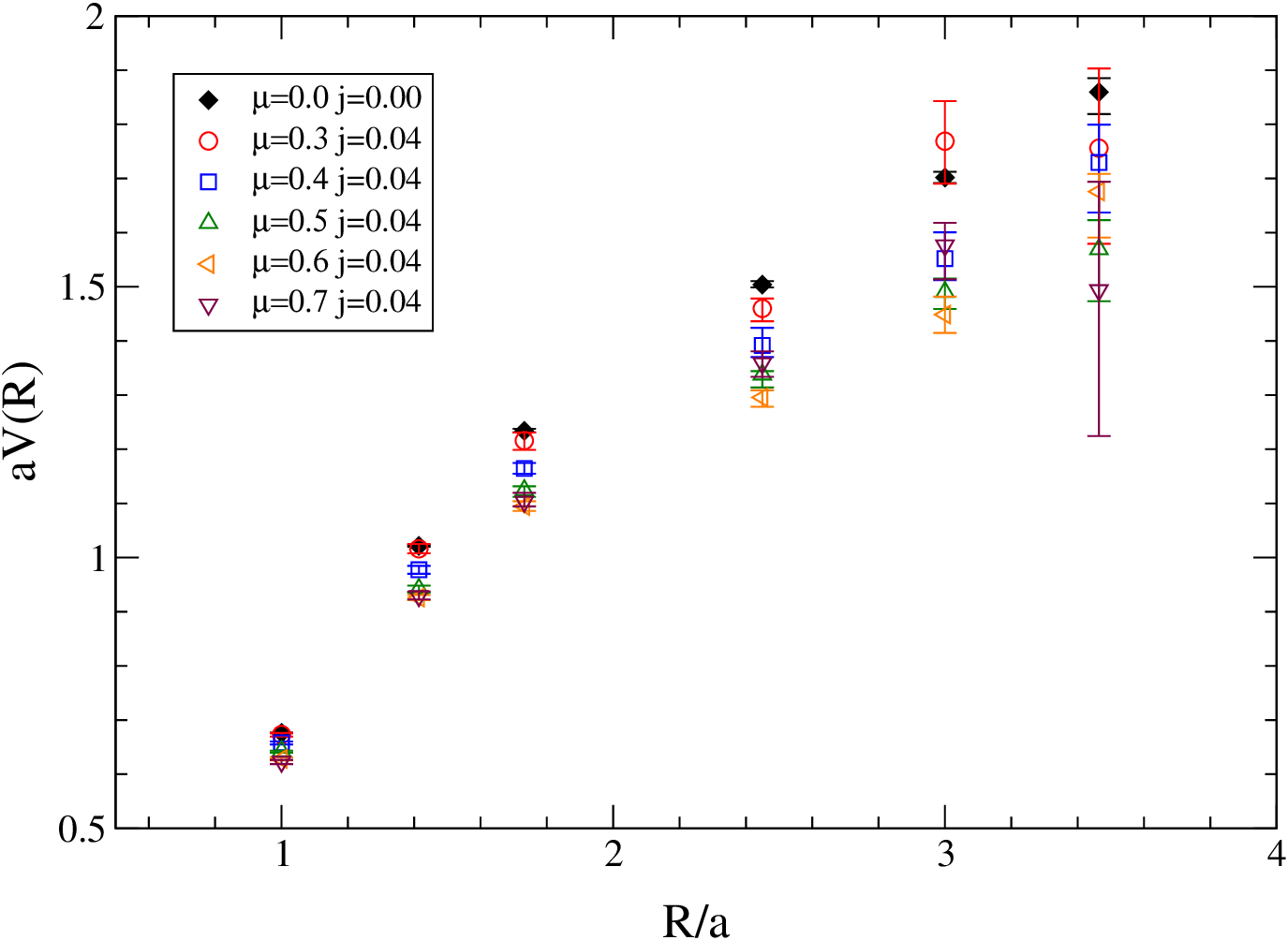}
\includegraphics*[width=0.45\textwidth]{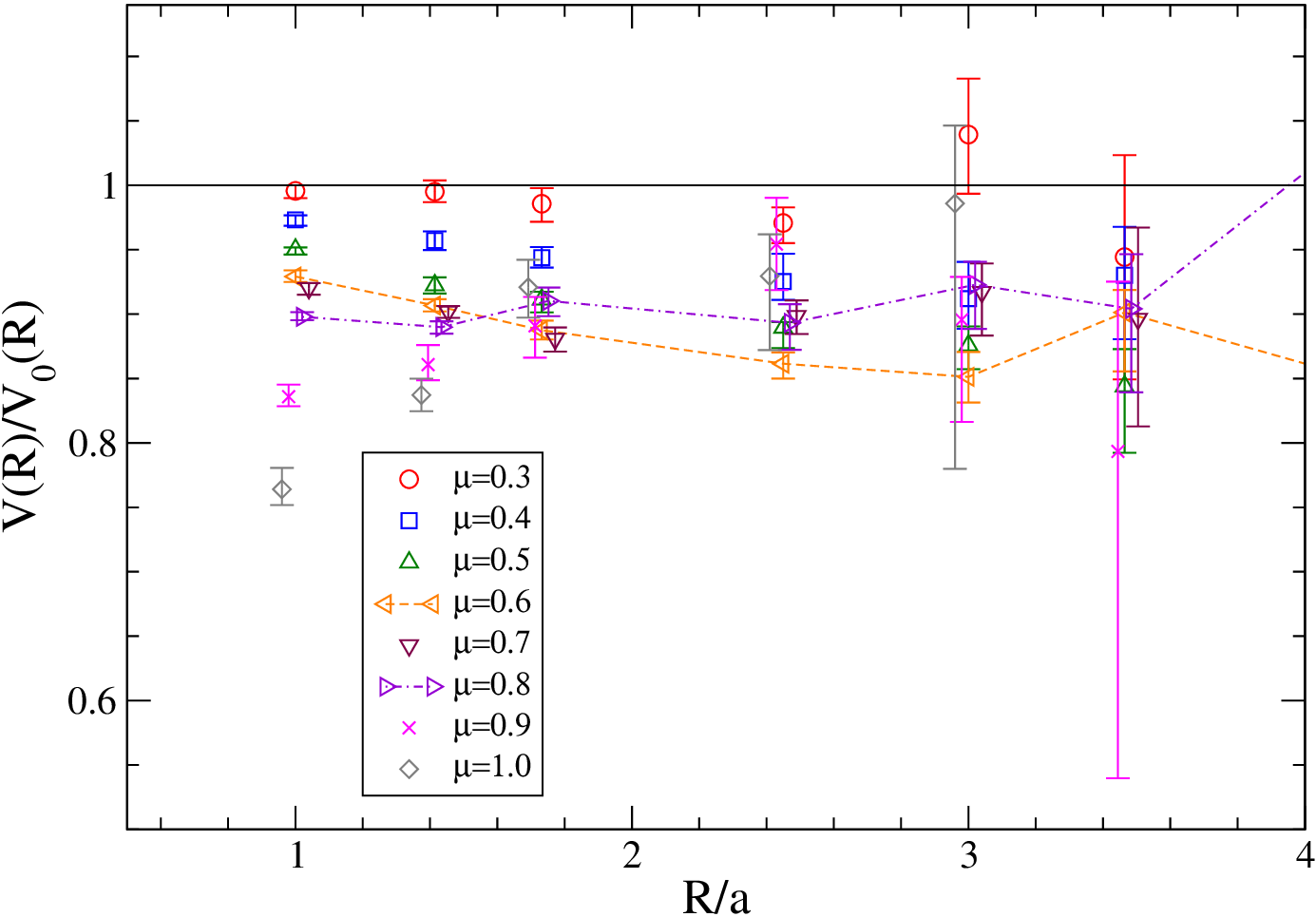}
\caption{The static quark potential for different values of the
  chemical potential $\mu$.  The right-hand plot shows the potential
  normalised to its value at $\mu=j=0$.}
\label{fig:potential}
\end{figure}

Finally, in figure \ref{fig:potential} we show the static quark
potential for selected values of $\mu$.  We see clear evidence of
screening for $a\mu\geq0.4$.  However, for $a\mu\gtrsim0.7$ a new
pattern appears to emerge, in that the short-distance potential is
strongly modified (and suppressed) while the long-distance screening,
if anything, is reversed.  This is most clearly seen in the right-hand
plot of fig.~\ref{fig:potential}, where the deviation of the potential
from that at $\mu=j=0$ is shown. The source of this behaviour is
subject of ongoing investigation.

\section{Discussion}

We have studied the $\mu$-dependence of local observables and the
static quark potential for 2-colour QCD with a quark mass well beyond
the expected range of validity for chiral perturbation theory.  We
find clear evidence of two separate (onset and deconfining)
transitions, and strong indications that a Fermi surface is formed,
leading to the formation of a BCS condensate at high densities.  In
this sense, the system is more QCD-like than 2-colour QCD at lighter
quark masses, where BEC dominates.

The simulations have been carried out with a fixed, non-zero diquark
source $j$.  We are currently performing simulations with a second,
smaller diquark source to enable an extrapolation to $j=0$.  This will
lay the grounds for drawing more definite conclusions regarding the
nature of the system in the different regions.

We are also studying the electric and magnetic gluon propagator,
which may provide useful input into the gap equation at high
densities, and are planning to study the hadron spectrum,
focusing in particular on the fate of the vector meson, which in a
previous study \cite{Muroya:2002ry} was found to become lighter at
higher density.

\section*{Acknowledgments}

This work has been supported by the IRCSET Embark Initiative award
SC/03/393Y.  SK thanks PPARC for support during his visit to Swansea
in 2004/05.  SJH was supported by a PPARC Senior Research
Fellowship. We have benefited from discussions with Kim Splittorff and
Jimmy Juge, and give our warm thanks to Shinji Ejiri and Luigi
Scorzato for their participation in the early stages of this project.

\bibliography{lattice,density}

\begin{thebibliography}{10}

\bibitem{Kogut:2000ek}
J. Kogut et~al.,
\newblock Nucl. Phys. B582 (2000) 477 [hep-ph/0001171].

\bibitem{Aloisio:2000if}
R. Aloisio et~al.,
\newblock Phys. Lett. B493 (2000) 189 [hep-lat/0009034].

\bibitem{Hands:2000ei}
S. Hands et~al.,
\newblock Eur. Phys. J. C17 (2000) 285 [hep-lat/0006018].

\bibitem{Hands:2001ee}
S. Hands et~al.,
\newblock Eur. Phys. J. C22 (2001) 451 [hep-lat/0109029].

\bibitem{Kogut:2001na}
J. Kogut et~al.,
\newblock Phys. Rev. D64 (2001) 094505 [hep-lat/0105026].

\bibitem{Muroya:2002ry}
S. Muroya, A. Nakamura and C. Nonaka,
\newblock Phys. Lett. B551 (2003) 305 [hep-lat/0211010].

\bibitem{Alles:2002st}
B. All{\'e}s et~al.,
\newblock hep-lat/0210039.

\bibitem{Skullerud:2003yc}
J.I. Skullerud et~al.,
\newblock Prog. Theor. Phys. Suppl. 153 (2004) 60 [hep-lat/0312002].

\bibitem{Hands:2005yq}
S. Hands, S. Kim and J.I. Skullerud,
\newblock PoS (LAT2005) (2005) 149 [hep-lat/0508027].

\bibitem{Skullerud:2004int}
J.I. Skullerud,
\newblock Simulating {SU}(2)-{QCD} with {W}ilson fermions, 2004,\\
\newblock
  http://mocha.phys.washington.edu/\~{}int\_talk/WorkShops/int\_04\_1/People/S%
kullerud\_J/.

\end{thebibliography}

\end{document}